\documentclass[a4paper,10pt,twoside]{article}
\usepackage{amsmath}
\usepackage{amssymb}
\usepackage{amsthm}
\usepackage{graphicx}

\theoremstyle{definition}

\theoremstyle{remark}

\newcommand{\bk}{{\bf k}}
\newcommand{\ba}{{\bf a}}
\newcommand{\br}{{\bf r}}

\newcommand{\bR}{{\bf R}}

\newcommand{\bX}{{\bf X}}
\newcommand{\bx}{{\bf x}}
\newcommand{\bz}{{\bf z}}
\newcommand{\by}{{\bf y}}
\newcommand{\bp}{{\bf p}}
\newcommand{\bu}{{\bf u}}
\newcommand{\bq}{{\bf q}}
\newcommand{\bQ}{{\bf Q}}
\newcommand{\bC}{{\bf C}}
\newcommand{\hk}{{\hat{\bf k}}}

\newcommand{\hx}{{\hat{\bf x}}}

\newcommand{\hz}{{\hat{\bf z}}}

\newcommand{\be}{\begin{equation}}
\newcommand{\ee}{\end{equation}}
\newcommand{\bay}{\begin{eqnarray}}
\newcommand{\eay}{\end{eqnarray}}
 \author{Koptelov Y.Y., Levin S.B.}
 \title{On the leading term of asymptotics of the $n$ like-charged quantum particles scattering problem solution. }

\begin{document}
 \addtolength{\hoffset}{-3.1cm}
  \addtolength{\voffset}{-3cm}

\maketitle

\abstract{An ansatz describing in terms of formal asymptotic decompositions a leading term of asymptotics of the $n$ three-dimensional like-charged quantum particles scattering problem solution is suggested. The description of the solution in those asymptotic configurations in which it was known earlier (for example, $n=3$), coincides with the previously known constructions \cite{BBK,z1,am92,BL2,KL}. It is shown that the Schredinger equation discrepancy for the suggested ansatz decreases faster than the potential uniformly in all angle variables at infinity in configuration space. An assumption is made about the structure of the leading term of the asymptotics of the scattering problem solution related to the $n$ three-dimensional quantum particles interacting by a broad class of slowly decreasing pair potentials. }

\section{Introduction}

It is well known that the leading term of the asymptotics of the $n$ quantum particles interacting by repulsive short-range pair potentials scattering problem solution represents a plane wave
$$
\Psi_0\ \sim \ e^{i\langle\bQ,\bX\rangle}, \ \bQ,\bX\in\bR^{d(n-1)},
$$
where $d$ -- is a particle's dimension. Whereas in the case when pair potentials are slowly decreasing (for example, the Coulomb ones), even the case $d=3,\ n=3$ represents a difficult problem. Nevertheless, for the three charged particles system a so called BBK-approximation has been known since the middle of the last century. It describes the leading term of the asymptotic of the scattering problem for some configurations, that is to say for such asymptotic configurations when all particles are well separated
$$
\Psi^{BBK}_c\ \sim \ e^{i\langle\bQ,\bX\rangle}\Phi_1(\bx_1,\bk_1)
\Phi_2(\bx_2,\bk_2)\Phi_3(\bx_3,\bk_3),\ \ \bQ,\bX\in\bR^6,\ \ \bx_j,\bk_j\in\bR^3,\ j=1,2,3.
$$
Here
\be
\Phi_j(\bx_j,\bk_j)=\Phi\left(-i\eta_j,1,i(|\bk_j||\bx_j|-\langle\bk_j,\bx_j\rangle)\right),\ \
\ j=1,2,3
\label{phi-def}
\ee
is an explicit hypergeometric function \cite{GR}, $\ \eta_j,\ j=1,2,3$ - is  a Sommerfeld parameter. This approximation was studied in \cite{BBK}, see also \cite{MF}, though it was used also earlier \cite{z1,z2}.
It is clear that such an asymptotic description was not complete, as the asymptotics of the scattering problem solution in the domains allowing the finite distances in particles pairs was missing.
This essential gap was partially filled by the works of Alt and Mukhamedzhanov \cite{am92}.
However, the methods used in these works did not allow to describe the asymptotics of the solution in the vicinities of forward scattering directions, uniformly in angle variables in all configuration space. It is required for example for solving corresponding boundary problems. Thus a necessity to develop new methods appeared.

One of the methods to describe the asymptotics of the three-dimensional charged quantum particles with repulsive pair potentials scattering problem solution has been developed in the recent years within the framework of the approach based on the analogy of the scattering problem with diffraction problem of the wave on a system of infinite semitransparent "screens" with vicinities (see \cite{BL1,BL2}
and references in the works). These "screens" represent a geometrical places of points in the configuration space of a many-particle system, in which the particles in pair subsystems coincide. Such an approach to the scattering problem allows to attain uniform asymptotics of the eigenfunctions of the continuous spectrum at infinity in configuration space on the basis of the heuristic constructions. One of the construction criteria is the discrepancy decrease velocity of the heuristically found Schredinger equation solution. In the work
\cite{KL} the approach developed in \cite{BL1,BL2} for a three-body system was generalized for a system of $n$ particles with possible finite distances in pairs which are all well separated from each other. In this work a uniform in all angle variables in configuration space asymptotics  of the $n$ like-charged quantum particles scattering problem is constructed on the level of formal asymptotic
decompositions.

In other words, an ansatz describing the leading term of the asymptptics of the $n$ like-charged quantum particles scattering problem solution is offered. An asymptotic description of the solution in those cases
in which it was known previously (for example, in $n=3$) coincides with the earlier known constructions \cite{BBK,z1,am92,BL2,KL}. It is shown  that the Schredinger equation discrepancy for the suggested ansatz decreases faster than the potential in all angle variables at infinity in configuration space.

\section{Statement of the problem }

We are studying a system of $n$ three-dimensional  equal mass particles interacting by equal pair Coulomb potentials.  The assumption of the equality of masses and potentials is introduced only to simplify the presentation. These limitations can be easily eliminated.

The initial configuration space of the system is  $\bR^{3n}$. Stopping the center of mass motion, we arrive at the system on configuration space
$$
\Gamma=\{\br:\, \br\in\bR^{3n},\,\br=\{\br_1,\br_2,...,\br_n\},\ \sum_{j=1}^{n}\br_j=0\}.
$$
On $\Gamma$ there is a scalar product $\langle\br,\br'\rangle$, induced by the scalar product
on $\bR^{3n}$. The system at $\Gamma$ is described by the equation
\be
H\Psi=E\Psi,\ \ \ \Psi=\Psi(\br)\in\bC,\ \ \br\in\Gamma,
\label{eq0}
\ee
\be
H=-\frac12\Delta_\br+V(\br), \quad V(\br) =\sum_{i,j=1;i<j}^n v(\br_i-\br_j),\ \ \ \br_l\in\bR^3, \ \ l=1,2,...,n.
\label{eq1}
\ee
Here $\Delta_\br$ -- is the Laplace operator on $\Gamma$.

We now proceed to Jacobi coordinates on $\Gamma$, keeping to the following scheme (see, for example, \cite{MF}).
Let $\omega_j$ - be a certain subsystem consisting of $j$ particles, then its' center of mass can be found:
\begin{equation}
\mathbf{w}_{\omega_j} = \frac{1}{j}\sum^{j}_{i=1}\mathbf{r}_{k_i},
\end{equation}
where $k_i$ - indexes the particles.
The coordinate $j+1$ of the particle relative to the subsystem $\omega_j$ looks as follows
\begin{equation}
\mathbf{x}_{\omega_jk_{j+1}}=
\left(\frac{2j}{j+1}\right)^{\frac{1}{2}}(\mathbf{w}_{\omega_j}-\mathbf{r}_{k_{j+1}}).
\label{jacobi}
\end{equation}
Starting from a certain particle $(j=1)$ and fixing the order of the inclusion of particles into a subsystem, we introduce on $\Gamma$ a set of $n-1$ orthogonal coordinates - Jacobi basis.

Alongside with the thus introduced coordinates $\mathbf{x}_{\omega_jk_{j+1}}\in {\bR^3}$
$(j=1,2,...,n-1)$ we will consider momenta dual to them according to Fourier transform:
$\bk_{\omega_jk_{j+1}}\in {\bR^3}$.

Note that the following relations are valid:
\be
\bx_{\omega_1m_{2}}=\sum_{j=1}^{n-1}\zeta^{(l)}_{mj}\bx_{\omega_jl_{j+1}},\ \ \ m=1,2,...,n-1,
\label{xref}
\ee
\be
\bk_{\omega_1m_{2}}=\sum_{j=1}^{n-1}\zeta^{(l)}_{mj}\bk_{\omega_jl_{j+1}},\ \ \ m=1,2,...,n-1,
\label{kref}
\ee
where the index $l$ is fixing the corresponding Jacobi basis, while the parameters $\zeta^{(l)}_{mj}$ --
are the transformation coefficients between different Jacobi bases.

The relation below is valid too:
\be
\sum_{j=1}^{n-1}(\zeta^{(l)}_{mj})^2=1.
\label{koeff}
\ee

We will assume that
$v(\bx_{\beta_1 l_2})=\frac{a_0}{|\bx_{\beta_1 l_2}|},\ \ a_0>0$, where
$\bx_{\beta_1 l_2}$ defines a relative coordinate in an arbitrary pair subsystem in accordance
with the Jacobi basis fixed by the parameter $\beta$.

We will finally re-write the Schredinger equation (\ref{eq0}) in the Jacobi coordinates
\be
H\Psi=E\Psi,\ \ \ \Psi=\Psi(\bX,\bQ)\in\bC,\ \ \bX\in\Gamma,\ \ \ E=Q^2,
\label{eq00}
\ee
\be
H=-\Delta_\bX+V(\bX), \quad V(\bX) =\sum_{\beta=1}^{n(n-1)/2}\frac{a_0}{|\bx_{\beta_1 l_2}|},\ \ \ \bx_{\beta_1 l_2}\in\bR^3.
\label{eq11}
\ee
Here $\Delta_\bX$ -- is a Laplace operator on  $\Gamma$,
$$
\bX=(\bx_{\omega_1\gamma_2},\bx_{\omega_2\gamma_3},...,\bx_{\omega_{n-1}\gamma_n})^t,\ \ \
\bQ=(\bk_{\omega_1\gamma_2},\bk_{\omega_2\gamma_3},...,\bk_{\omega_{n-1}\gamma_n})^t,
$$
where the parameter $\gamma$ is fixing a certain set of Jacobi coordinates, connected by the construction with a certain two-particle subsystem.

We can also introduce more complicated Jacobi bases connected with several many-particle subsystems.
Let there be $l$ subsystems in the $n$ particles system, each subsystem consisting of $m_j$ particles,
$\ j=1,2,...,l$.
On each of such subsystems a Jacobi basis is introduced, consisting of $m_j-1$ coordinates $\by_i^{(j)}$, $\ j=1,2,...,l$, $\ i=1,2,...,m_j-1$ in the way described above. After that a complementary Jacobi basis is introduced for a system consisting of $n-\sum_{j=1}^l m_j$ particles and $l$ "quasi particles". The mass of each such "quasi particle" is equal to a summation of all masses of the particles included into the subsystem, while the coordinate of the "quasi particle" coincides with the center of mass of the subsystem. We will call the coordinates of the complementary Jacobi basis $\bz_i$,
$\ i=1,2,...,n-\sum_{j=1}^l m_j+l-1$.
The complex of the $l+1$ bases thus arisen $\left\{\{\by^{(1)}\},...,\{\by^{(l)}\},\{\bz\}\right\}$
forms a complete Jacobi basis in the system of $n$ bodies.

We will finally note that all the Jacobi bases constructed in a variety of ways in the complete $n$ bodies system are connected with each other by rotation transformations.

\section{Result formulation}

We will outline the structure of the ansatz describing the leading term of asympotics at infinity in configuration space of the scattering problem solution in the system of $n$ like-charged quantum particles. An essential assumption is as follows. In the further considerations we assume that we know the solutions (not asymptotic but complete) of the scattering problems in all subsystems of the complete $n$ particles system.

We will consider first a configuration when all particles are well separated, i.e.
$$
|\bx_j|\rightarrow\infty,\ \ j=1,2,...,\frac{n(n-1)}{2}.
$$
Here $\bx_j,\ j=1,2,...,\frac{n(n-1)}{2}$ -- are pair coordinates.

In this case the asymptotic looks as follows:
\be
\Psi^{BBK}_c\ \sim \ e^{i\langle\bQ,\bX\rangle}\mathop{\prod}\limits_{j=1}^{\frac{n(n-1)}{2}}
\Phi_j(\bx_j,\bk_j),\ \
\bX=\left(\begin{tabular}{c}
      $\by_1$\\
      $\by_2$\\
      $\dots$\\
      $\by_{n-1}$\\
\end{tabular}\right),\ \ \ \
\bQ=\left(\begin{tabular}{c}
      $\bp_1$\\
      $\bp_2$\\
      $\dots$\\
      $\bp_{n-1}$\\
\end{tabular}\right).
\label{res-00}
\ee
As well as above, here  $\Phi_j(\bx_j,\bk_j)=\Phi\left(-i\eta_j,1,i(|\bk_j||\bx_j|-\langle\bk_j,\bx_j\rangle)\right)$ -
is an explicit hypergeometrical function \cite{GR}, $\ \eta_j,\ j=1,2,3$ - is a Sommerfeld parameter,
$\by_i,\ i=1,2,...,n-1$ -- are the coordinates in the Jacobi basis, $\bp_i,\ i=1,2,...,n-1$ -- are the momenta conjugated to them.

In the work \cite{KL} it was shown that the Schredinger equation discrepancy for such an approximation decreases at infinity in configuration space uniformly in all angle variables faster than the potential.

We will assume now that with the hyperradius of the system
$R=\left(\mathop{\sum}\limits_{j=1}^{\frac{n(n-1)}{2}}|\bx_j|^2\right)^2$ going to infinity, some pair coordinates turn out to be limited, i.e. for a certain subset of whole numbers $\sigma\in\{1,2,...,\frac{n(n-1)}{2}\}$ the following relation is fulfilled
\be
|\bx_j|\le \Omega <\infty,\ \ j\in\sigma,\ \ \ R\rightarrow\infty.
\label{cond}
\ee

We will assume that the considered $n$ particles system contains $l$ subsystems, each consisting
of $m_j,\ j=1,2,...,l$ particles. Here the condition  (\ref{cond}) is met for all pair coordinates in the subsystems and only for them. In this sense we will call such subsystems the
"clusters".

We will introduce into our consideration the functions
$$
\chi_j(\bX_j,\bQ_j),\ \ \ \
\bX_j=\left(\begin{tabular}{c}
      $\by^{(j)}_1$\\
      $\by^{(j)}_2$\\
      $\dots$\\
      $\by^{(j)}_{m_j-1}$\\
\end{tabular}\right),\ \ \ \
\bQ_j=\left(\begin{tabular}{c}
      $\bp^{(j)}_1$\\
      $\bp^{(j)}_2$\\
      $\dots$\\
      $\bp^{(j)}_{m_j-1}$\\
\end{tabular}\right),\ \ \ j=1,2,...,l\ -
$$
continuous spectrum eigenfunctions of the isolated "clusters" energy operators. These functions satisfy the Schredinger equation of the form
\be
-\mathop{\sum}\limits_{\beta=1}^{m_j-1}\Delta_{\by^{(j)}_\beta}\chi_j+
\mathop{\sum}\limits_{\alpha=1}^{\frac{m_j(m_j-1)}{2}}\frac{a_0}{|\bx^{(j)}_\alpha|}\chi_j=
\mathop{\sum}\limits_{\beta=1}^{m_j-1}|\bp^{(j)}_\beta|^2\chi_j.
\label{eqpart}
\ee

For such configurations the asymptotics looks as follows:
\be
\Psi^{BBK}_c\ \sim \ e^{i\langle\bQ_0,\bX_0\rangle}
\mathop{\prod}\limits_{j=1}^{l}\chi_j(\bX_j,\bQ_j)
\mathop{\prod}\limits_{i=M+1}^{\frac{n(n-1)}{2}}\tilde{\Phi}_i(\tilde{\bx}_i,\bk_i),
\label{result-0}
\ee
$$
\bX_0=\left(\begin{tabular}{c}
      $\bz_{N+1}$\\
      $\bz_{N+2}$\\
      $\dots$\\
      $\bz_{n-1}$\\
\end{tabular}\right),\ \ \ \
\bQ_0=\left(\begin{tabular}{c}
      $\bq_{N+1}$\\
      $\bq_{N+2}$\\
      $\dots$\\
      $\bq_{n-1}$\\
\end{tabular}\right),\ \ \ N=\mathop{\sum}\limits_{j=1}^l(m_j-1),\ \ \ M=\mathop{\sum}\limits_{j=1}^l\frac{m_j(m_j-1)}{2}.
$$
We used here the definition
\be
\tilde{\Phi}_j(\tilde{\bx}_j,\bk_j)\equiv
\Phi\left(-i\eta_j,1,i(|\bk_j||\tilde{\bx}_j|-\langle\bk_j,\tilde{\bx}_j\rangle)\right).
\label{def-phi-0}
\ee

Note that the function  $\tilde{\Phi}_i(\tilde{\bx}_i,\bk_i),\ \ i=M+1,M+2,...,\frac{n(n-1)}{2}$
differs from the expressions defined above in the equation (\ref{phi-def}) by the transformation of the coordinate $\bx_i$:
\be
\bx_i\ \rightarrow \tilde{\bx}_i,\ \ \ i=M+1,M+2,...,\frac{n(n-1)}{2}.
\label{trans-1}
\ee
The essence of this transformation is a key place in the ansatz construction. We will write first the expression $\bx_i$ in the terms of the Jacobi basis constructed above in the considered $n$ particles system:
\be
\bx_i=\mathop{\sum}\limits_{j=1}^{l}\mathop{\sum}\limits_{\nu=1}^{m_j-1}
\zeta_{i\nu}^{(j)}\by^{(j)}_\nu\ +\ \mathop{\sum}\limits_{\nu=N+1}^{n-1}\zeta_{i\nu}^{(0)}\bz_\nu.
\label{trans-2}
\ee
Note that by the construction all the coordinates
$\by^{(j)}_\nu,\ \ \ \ j=1,2,...,l;\ \ \nu=1,2,...,m_j-1$
of the Jacobi basis are finite and satisfy the condition (\ref{cond}):
$$
|\by^{(j)}_\nu|\le\Omega<\infty.
$$
On the opposite, all the coordinates $\bz_\nu,\ \ \ \ \nu=N+1,N+2,...,n-1$ are infinite ones.
The transformation (\ref{trans-1}) consists in the following substitution in the equation (\ref{trans-2}):
\be
\by^{(j)}_\nu\ \rightarrow\ u^{(j)}_\nu\ =\
-i\frac{\nabla_{\bp_\nu^{(j)}}\chi_j(\bX_j,\bQ_j)}{\chi_j(\bX_j,\bQ_j)},\ \ j=1,2,...,l;\ \ \nu=1,2,...,m_j-1.
\label{trans-3}
\ee
Thus the expression $\tilde{\bx}_i,\ \ i=M+1,M+2,...,\frac{n(n-1)}{2}$ (\ref{trans-1})
looks as follows:
\be
\tilde{\bx}_i=\mathop{\sum}\limits_{j=1}^{l}\mathop{\sum}\limits_{\nu=1}^{m_j-1}
\zeta_{i\nu}^{(j)}u^{(j)}_\nu\ +\ \mathop{\sum}\limits_{\nu=N+1}^{n-1}\zeta_{i\nu}^{(0)}\bz_\nu.
\label{trans-4}
\ee

The expressions (\ref{result-0})-(\ref{trans-4}) fully define the structure of the ansatz suggested.

Note that with  $|\bx^{(j)}_i|\rightarrow\infty,\ \ i=1,2,...,\frac{m_j(m_j-1)}{2}$
$$
\chi_j(\bX_j,\bQ_j)\rightarrow e^{i\langle\bX_j,\bQ_j\rangle}
\mathop{\prod}\limits_{i=1}^{\frac{m_j(m_j-1)}{2}}\Phi_i(\bx^{(j)}_i,\bk^{(j)}_i).
$$

Thus it is clear that if all pair coordinates in the subsystem go to infinity, the expression (\ref{result-0}) goes into the expression (\ref{res-00}) with an accuracy up to the value which does not influence the discrepancy decrease velocity in the main order.

\section{ Verification of the constructions suggested}

We will study the Schredinger equation discrepancy behaviour for the suggested constructions.
Let us consider first a "cluster" of $n-1$ particles and one isolated particle.

\subsection{Notations system }

Let the Jacobi system of coordinates connected with $n-1$ particles in "the cluster" and arising after a separation of the center of mass, looks as follows:
$$
\by_j,\ \ j=2,3,...,n-1.
$$

 In this connection we assume that the absolute values of all Jacobi coordinates in the "cluster" are limited:
$$
|\by_j|\leq \Omega < \infty,\ \ \ j=2,3,...,n-1.
$$
The Jacobi coordinate $\bz_1$, describing the dynamics of the separated particle relative to the centre of mass of the "cluster"\mbox{,} satisfies the condition
$$
|\bz_1|\gg \Omega.
$$

Let all pair coordinates in the $n$ particles system $\bx_\alpha$ be numbered by the index $\alpha$,
$$
\alpha=1,2,...,\frac{n(n-1)}{2}.
$$

While for the pair subsystems not containing a well separated particle
\be
\bx_\alpha=\zeta_{\alpha 2}\by_2+...+\zeta_{\alpha n-1}\by_{n-1},\ \ \ \ \
\alpha=1,2,...,\frac{(n-1)(n-2)}{2}.
\label{e1}
\ee

For subsystems containing a well separated particle
\be
\bx_\alpha=\zeta_{\alpha 1}\bz_1+\zeta_{\alpha 2}\by_2+...+\zeta_{\alpha n-1}\by_{n-1},\ \ \ \ \
\alpha=\frac{(n-1)(n-2)}{2}+1,...,\frac{n(n-1)}{2}.
\label{e2}
\ee
Note that the conditions of the rotation transformation for various Jacobi bases are met too.
$$
\mathop{\sum}\limits_{j=1}^{n-1}\zeta_{\alpha j}^2=1.
$$

For the momenta conjugated to the Jacobi coordinates the relations are fulfilled analogous to
(\ref{e1}),(\ref{e2}):
\be
\bk_\alpha=\zeta_{\alpha 2}\bp_2+...+\zeta_{\alpha n-1}\bp_{n-1},\ \ \ \ \
\alpha=1,2,...,\frac{(n-1)(n-2)}{2}.
\label{k1}
\ee
\be
\bk_\alpha=\zeta_{\alpha 1}\bq_1+\zeta_{\alpha 2}\bp_2+...+\zeta_{\alpha n-1}\bp_{n-1},\ \ \ \ \
\alpha=\frac{(n-1)(n-2)}{2}+1,...,\frac{n(n-1)}{2}.
\label{k2}
\ee

\subsection{Ansatz structure for a configuration: one well separated particle and a "cluster"
consisting of $n-1$ particles. }

The ansatz structure in the case considered looks as follows:
\be
\Psi(\bX,\bQ)=e^{i\langle\bq_1,\bz_1\rangle}
\chi(\bX,\bQ)
\mathop{\prod}\limits_{\alpha\in \{M_0\}}\tilde{\Phi}_\alpha(\tilde{\bx}_\alpha,\bk_\alpha),
\label{anz}
\ee
$$
\bX=\left(\begin{tabular}{c}
      $\bz_1$\\
      $\by_2$\\
      $\dots$\\
      $\by_{n-1}$\\
\end{tabular}\right),\ \ \ \
\bQ=\left(\begin{tabular}{c}
      $\bq_1$\\
      $\bp_2$\\
      $\dots$\\
      $\bp_{n-1}$\\
\end{tabular}\right).
$$

Here the function $\chi$ is a scattering problem solution in the system of $n-1$ particles included into the "cluster"
\mbox{,}
and satisfies the Schredinger equation
\be
-\mathop{\sum}\limits_{\beta=2}^{n-1}\Delta_{\by_\beta}\chi+
V_1\chi=E_1\chi.
\label{shr-c}
\ee

Here we use the notations
\be
V_1=\mathop{\sum}\limits_{\alpha\in \{M_1\}}\frac{a_0}{|\bx_\alpha|},\ \ \ \
E_1=\mathop{\sum}\limits_{\beta=2}^{n-1}\bp_\beta^2,
\label{claster}
\ee
$$
\{M_0\}=\left\{1+\frac{(n-1)(n-2)}{2},2+\frac{(n-1)(n-2)}{2},...,n-1+\frac{(n-1)(n-2)}{2}\right\},
$$
$$
\{M_1\}=\left\{1,2,...,\frac{(n-1)(n-2)}{2}\right\}.
$$
Functions $\tilde{\Phi}_\alpha(\tilde{\bx}_\alpha,\bk_\alpha)$ represent a modification of the explicit hypergeometrical function
$$
\Phi\left(-i\eta_\alpha,1,i(|\bx_\alpha||\bk_\alpha|-\langle\bx_\alpha,\bk_\alpha\rangle)\right),\ \ \
 \alpha\in\{M_0\}.
 $$
with each vector coordinate in the third argument
$$
\bx_\alpha=\zeta_{\alpha 1}\bz_1+\zeta_{\alpha 2}\by_2+...+\zeta_{\alpha n-1}\by_{n-1},\ \ \ \ \
\alpha=\frac{(n-1)(n-2)}{2}+1,...,\frac{n(n-1)}{2}
$$
undergoing a change
$$
\bx_\alpha\ \rightarrow\ \tilde{\bx}_\alpha,
$$
where
$$
\tilde{\bx}_\alpha=\zeta_{\alpha 1}\bz_1+\zeta_{\alpha 2}\bu_2+...
+\zeta_{\alpha n-1}\bu_{n-1},\ \ \ \ \
\alpha=\frac{(n-1)(n-2)}{2}+1,...,\frac{n(n-1)}{2},
$$
\be
\bu_j\ \equiv\ -i\frac{\nabla_{\bp_j}\chi}{\chi},\ \ \ \ j=2,3,...,n-1.
\label{def0}
\ee

We remind that the function $\chi$ satisfies the equation (\ref{shr-c})and depends on the above introduced set of Jacobi variables $\by_2,\by_3,...,\by_{n-1}$,
\be
\chi=\chi(\by_2,\by_3,...,\by_{n-1};\bp_2,\bp_3,...,\bp_{n-1}).
\label{chi1}
\ee

The structure of the presented modification is dictated by a method developed earlier in the works
\cite{BL1}, \cite{BL2}, \cite{KL}, \cite{K}.

Note too that as was shown in the work \cite{KL} for an analogous case, the asymptotic
transition at $\by_\alpha\ \rightarrow\ \infty,\ \ \ \ \alpha\in\{M_0\}$ is valid:
$$
\Psi(\bX,\bQ)\ \rightarrow\ e^{i\langle\bQ,\bX\rangle}
\mathop{\prod}\limits_{\alpha=1}^{\frac{n(n-1)}{2}}\Phi_\alpha.
$$

A correction to the asymptotics contains only the terms, the discrepancy of which in the Schredinger equation decreases necessarily faster than the potential.

\subsection{The construction of the Schredinger equation discrepancy and the evaluations of the velocity of its' decrease at infinity in configuration space.}

We will show now that the discrepancy of the expression described in (\ref{anz}) decreases
faster than the potential uniformly in all angle variables in configuration space. Here and below all evaluations are considered beyond narrow vicinities of the directions
$\langle\hx_\alpha,\hk_\alpha\rangle=1,\ \ \ \alpha\in\{M_0\}$. We can however show that also within these narrow vicinities the discrepancy decrease will be faster than the potential, even though we cannot construct for that the estimation from below. To the subject of this special case a separate article will be dedicated.

\subsubsection{ Intermediate estimations}

We will give a number of intermediate estimations which will be required for further computations:
\be
|\bx_\alpha|=|\zeta_{\alpha 1}||\bz_1|+\varepsilon_{\alpha 1}\mathop{\sum}\limits_{\beta=2}^{n-1}
\langle\hz_1,\zeta_{\alpha \beta}\by_\beta\rangle
+ O\left(\frac{1}{|\bz_1|}\right),\ \ \ \ \alpha\in\{M_0\},
\label{est1}
\ee
where a following notation is introduced $\varepsilon_{\alpha 1}\equiv sign(\zeta_{\alpha 1})$.

The next estimation looks as follows:
\be
|\bx_\alpha||\bk_\alpha|-\langle\bk_\alpha,\bx_\alpha\rangle=
|\zeta_{\alpha 1}|\left(|\bk_\alpha||\bz_1|-
\varepsilon_{\alpha 1}\langle\bk_\alpha,\bz_1\rangle\right)+
\varepsilon_{\alpha 1}|\bk_\alpha|\mathop{\sum}\limits_{\beta=2}^{n-1}
\langle\hz_1-\varepsilon_{\alpha 1}\hk_\alpha,\zeta_{\alpha\beta}\by_{\beta}\rangle
+O\left(\frac{1}{|\bz_1|}\right).
\label{est2}
\ee

For further computations we will need a notation $\bu_j,\ \ j=2,3,...,n-1$, introduced above in the equation (\ref{def0}):
\be
\nabla_{\by_\beta}\tilde{\Phi}_\alpha=i{\tilde{\Phi}}'_\alpha
\varepsilon_{\alpha 1}|\bk_\alpha|
\mathop{\sum}\limits_{\omega=2}^{n-1}\nabla_{\by_\beta}
\langle\hz_1-\varepsilon_{\alpha 1}\hk_\alpha,\zeta_{\alpha\omega}\bu_\omega\rangle,
\label{ny}
\ee
\be
\Delta_{\by_\beta}\tilde{\Phi}_\alpha=-{\tilde{\Phi}}^{''}_\alpha |\bk_\alpha|^2
\left|\mathop{\sum}\limits_{\omega=2}^{n-1}\nabla_{\by_\beta}\langle\hz_1-\varepsilon_{\alpha 1}
\hk_\alpha,\zeta_{\alpha\omega}\bu_\omega\rangle\right|^2+
i|\bk_\alpha|\varepsilon_{\alpha 1}{\tilde{\Phi}}'_\alpha
\mathop{\sum}\limits_{\omega=2}^{n-1}\Delta_{\by_\beta}\langle\hz_1-\varepsilon_{\alpha 1}
\hk_\alpha,\zeta_{\alpha\omega}\bu_\omega\rangle,
\label{dy}
\ee
\be
\nabla_{\bz_1}\tilde{\Phi}_\alpha=i{\tilde{\Phi}}'_\alpha|\bk_\alpha||\zeta_{\alpha 1}|
\left(\hz_1-\varepsilon_{\alpha 1}\hk_\alpha\right)+O\left(\frac{1}{|\bz_1|^2}\right),
\label{nz}
\ee
\be
\Delta_{\bz_1}\tilde{\Phi}_\alpha=O\left(\frac{1}{|\bz_1|^2}\right).
\label{dz}
\ee

We will apply now a Laplace operator to the full solution
$\Psi(\bX,\bQ)$:
\be
\Delta_{\bz_1}\Psi=-\bq_1^2e^{i\langle\bq_1,\bz_1\rangle}\chi
\mathop{\prod}\limits_{\alpha\in\{M_0\}}\tilde{\Phi}_\alpha\ \ +
\ \ 2ie^{i\langle\bq_1,\bz_1\rangle}\chi
\mathop{\sum}\limits_{\alpha\in\{M_0\}}\langle\bq_1,\nabla_{\bz_1}\tilde{\Phi}_\alpha\rangle
\mathop{\prod}\limits_{\gamma\in\{M_0\},\ \gamma\neq\alpha}\tilde{\Phi}_\gamma\ \ +
\label{dz1}
\ee
$$
+\ \ e^{i\langle\bq_1,\bz_1\rangle}\mathop{\sum}\limits_{\alpha\in\{M_0\}}\Delta_{\bz_1}
\tilde{\Phi}_\alpha
\mathop{\prod}\limits_{\gamma\in\{M_0\},\ \gamma\neq\alpha}\tilde{\Phi}_\gamma,
$$
\be
\Delta_{\by_\beta}\Psi=e^{i\langle\bq_1,\bz_1\rangle}\chi
\mathop{\sum}\limits_{\alpha\in\{M_0\}}\Delta_{\by_\beta}
\tilde{\Phi}_\alpha
\mathop{\prod}\limits_{\gamma\in\{M_0\},\ \gamma\neq\alpha}\tilde{\Phi}_\gamma\ \ +
\ \ 2e^{i\langle\bq_1,\bz_1\rangle}\chi
\mathop{\sum}\limits_{\gamma<\alpha\in\{M_0\}}\langle\nabla_{\by_\beta}\tilde{\Phi}_\gamma,
\nabla_{\by_\beta}\tilde{\Phi}_\alpha\rangle
\mathop{\prod}\limits_{\omega\in\{M_0\},\ \omega\neq\alpha,\ \omega\neq\gamma}\tilde{\Phi}_\omega\ \ +
\label{dyb}
\ee
$$
+\ \ 2e^{i\langle\bq_1,\bz_1\rangle}
\mathop{\sum}\limits_{\alpha\in\{M_0\}}\langle\nabla_{\by_\beta}\tilde{\Phi}_\alpha,
\nabla_{\by_\beta}\chi\rangle
\mathop{\prod}\limits_{\gamma\in\{M_0\},\ \gamma\neq\alpha}\tilde{\Phi}_\gamma\ \ +
\ \ e^{i\langle\bq_1,\bz_1\rangle}\Delta_{\by_\beta}\chi
\mathop{\prod}\limits_{\alpha\in\{M_0\}}\tilde{\Phi}_\alpha\ \ +\ \ O\left(\frac{1}{|\bz_1|^2}\right).
$$

\subsubsection{The discrepancy structure}

After the substitution of the expressions (\ref{ny})-(\ref{dz}) into the equation (\ref{dz1})-(\ref{dyb}) we can obtain an expression for the discrepancy
\be
S=(H-E)\Psi,\ \ \ H=-\Delta_{\bz_1}-\mathop{\sum}\limits_{\beta=2}^{n-1}\Delta_{\by_\beta}+
\mathop{\sum}\limits_{\alpha=1}^{\frac{n(n-1)}{2}}\frac{a_0}{|\bx_\alpha|}.
\label{disc}
\ee

Note that beyond the vicinities of special directions from the hypergeometrical equation for the function
$\tilde{\Phi}_\alpha$
$$
2\bk_\alpha^2(1-\langle\hk_\alpha,\hat{\tilde{\bx}}_\alpha\rangle){\tilde{\Phi}}^{''}_\alpha
+\frac{a_0}{|\tilde{\bx}_\alpha|}\tilde{\Phi}_\alpha-2i|\bk_\alpha|
\frac{1}{|\tilde{\bx}_\alpha|}{\tilde{\Phi}}^{'}_\alpha-
2\bk_\alpha^2(1-\langle\hk_\alpha,\hat{\tilde{\bx}}_\alpha\rangle){\tilde{\Phi}}^{'}_\alpha=0
$$
there follows with regard to(\ref{est1})-(\ref{est2}) a relation
$$
\frac{a_0}{|\bx_\alpha|}\tilde{\Phi}_\alpha=
2\bk_\alpha^2(1-\varepsilon_{\alpha 1}\langle\hz_1,\hk_\alpha\rangle){\tilde{\Phi}}^{'}_\alpha
+O\left(\frac{1}{|\bz_1|^2}\right).
$$

Keeping in the discrepancy only the terms decreasing at infinity in configuration space not faster than  the Coulomb potential, we obtain with regard to the previous evaluations
\be
S=2\mathop{\sum}\limits_{\alpha\in\{M_0\}}|\bk_\alpha||\zeta_{\alpha 1}|{\tilde{\Phi}}^{'}_\alpha
\langle\bq_1,\hz_1-\varepsilon_{\alpha 1}\hk_\alpha\rangle
\mathop{\prod}\limits_{\gamma\in\{M_0\},\ \gamma\neq\alpha}\tilde{\Phi}_\gamma\chi\ -\
\label{disc11}
\ee
$$
-\ i\mathop{\sum}\limits_{\alpha\in\{M_0\}}\varepsilon_{\alpha 1}|\bk_\alpha|{\tilde{\Phi}}^{'}_\alpha
\mathop{\sum}\limits_{\beta=2}^{n-1}\Delta_{\by_\beta}\langle\hz_1-\varepsilon_{\alpha 1}
\hk_\alpha,\mathop{\sum}\limits_{\omega=2}^{n-1}\zeta_{\alpha\omega}\bu_\omega\rangle
\mathop{\prod}\limits_{\gamma\in\{M_0\},\ \gamma\neq\alpha}\tilde{\Phi}_\gamma\chi\ -
$$
$$
-\ 2i\mathop{\sum}\limits_{\alpha\in\{M_0\}}\varepsilon_{\alpha 1}|\bk_\alpha|{\tilde{\Phi}}^{'}_\alpha
\mathop{\sum}\limits_{\beta=2}^{n-1}
\left\langle\nabla_{\by_\beta}\langle\hz_1-\varepsilon_{\alpha 1}
\hk_\alpha,\mathop{\sum}\limits_{\omega=2}^{n-1}\zeta_{\alpha\omega}\bu_\omega\rangle,
\nabla_{\by_\beta}\chi\right\rangle
\mathop{\prod}\limits_{\gamma\in\{M_0\},\ \gamma\neq\alpha}\tilde{\Phi}_\gamma\ +
$$
$$
+\ 2\mathop{\sum}\limits_{\alpha\in\{M_0\}}|\bk_\alpha|^2{\tilde{\Phi}}^{'}_\alpha
\left(1-\varepsilon_{\alpha 1}\langle\hz_1,\hk_\alpha\rangle\right)
\mathop{\prod}\limits_{\gamma\in\{M_0\},\ \gamma\neq\alpha}\tilde{\Phi}_\gamma\chi\ +\
O\left(\frac{1}{|\bz_1|^2}\right).
$$

We rewrite the equation (\ref{disc11}) for the discrepancy in a simpler way
\be
S=\mathop{\sum}\limits_{\alpha\in\{M_0\}}S_\alpha
{\tilde{\Phi}}^{'}_\alpha
\mathop{\prod}\limits_{\gamma\in\{M_0\},\ \gamma\neq\alpha}\tilde{\Phi}_\gamma\ +\
O\left(\frac{1}{|\bz_1|^2}\right),
\label{disca}
\ee
where the coefficient $S_\alpha$ looks as follows:
\be
S_\alpha\ =\ 2|\bk_\alpha||\zeta_{\alpha 1}|
\langle\bq_1,\hz_1-\varepsilon_{\alpha 1}\hk_\alpha\rangle\chi\ -\
\label{disc11-1}
\ee
$$
-\ i\varepsilon_{\alpha 1}|\bk_\alpha|
\mathop{\sum}\limits_{\beta=2}^{n-1}\Delta_{\by_\beta}\langle\hz_1-\varepsilon_{\alpha 1}
\hk_\alpha,\mathop{\sum}\limits_{\omega=2}^{n-1}\zeta_{\alpha\omega}\bu_\omega\rangle
\chi\ -
$$
$$
-\ 2i\varepsilon_{\alpha 1}|\bk_\alpha|
\mathop{\sum}\limits_{\beta=2}^{n-1}
\left\langle\nabla_{\by_\beta}\langle\hz_1-\varepsilon_{\alpha 1}
\hk_\alpha,\mathop{\sum}\limits_{\omega=2}^{n-1}\zeta_{\alpha\omega}\bu_\omega\rangle,
\nabla_{\by_\beta}\chi\right\rangle\ +
$$
$$
+\ 2|\bk_\alpha|^2
\left(1-\varepsilon_{\alpha 1}\langle\hz_1,\hk_\alpha\rangle\right)
\chi.
$$

Consider now a behaviour of a simpler object - a coefficient $S_\alpha$.
Our goal is to prove the following statement:
\be
S_\alpha=0,\ \ \ \alpha\in\{M_0\}.
\label{def11}
\ee

This statement is equivalent to the statement that full Schredinger equation discrepancy
 for the suggested scattering problem solution asymptotics of $n$ like-charged quantum
 particles decreases at infinity in configuration space faster than the Coulomb potential.

\subsubsection{Proof of the statement about the decrease velocity of Schredinger equation discrepancy.}

We will need an equation (\ref{k2}) quoted as follows:
\be
\bq_1=\frac{1}{\zeta_{\alpha 1}}\bk_\alpha-\frac{1}{\zeta_{\alpha 1}}
\mathop{\sum}\limits_{\omega=2}^{n-1}\zeta_{\alpha\omega}\bp_\omega,\ \ \ \ \alpha\in\{M_0\}.
\label{k2-1}
\ee

Here $\zeta_{\alpha 1}\neq 0$, as the equation for the conjugated (in the sense of Fourier transform) coordinates
(\ref{e2}) at $\alpha\in\{M_0\}$ necessarily contains a big parameter $\bz_1$ with a non-zero
coefficient.

We will substitute the equation (\ref{k2-1}) into the sum of the expressions in the first and
the fourth lines of the equation (\ref{disc11-1}):
\be
2|\bk_\alpha||\zeta_{\alpha 1}|
\langle\bq_1,\hz_1-\varepsilon_{\alpha 1}\hk_\alpha\rangle\chi\ +\
 2|\bk_\alpha|^2
\left(1-\varepsilon_{\alpha 1}\langle\hz_1,\hk_\alpha\rangle\right)
\chi\ =\
-2|\bk_\alpha|\varepsilon_{\alpha 1}\mathop{\sum}\limits_{\omega=2}^{n-1}\zeta_{\alpha\omega}
\langle\bp_\omega,\hz_1-\varepsilon_{\alpha 1}\hk_\alpha\rangle\chi.
\label{ring}
\ee

With regard to the equation (\ref{ring}) the expression for the coefficient  $S_\alpha$ (\ref{disc11-1})
is rewritten as
\be
S_\alpha=-|\bk_\alpha|\varepsilon_{\alpha 1}\mathop{\sum}\limits_{\omega=2}^{n-1}\zeta_{\alpha\omega}
\sigma_\omega^\alpha,
\label{ring1}
\ee
where $\sigma_\omega$ is expressed as follows:
\be
\sigma_\omega^\alpha=2\langle\bp_\omega,\ba_\alpha\rangle\chi\ +\
\mathop{\sum}\limits_{\beta=2}^{n-1}\Delta_{\by_\beta}\frac{g_{\alpha\omega}}{\chi}\chi\ +\
2\mathop{\sum}\limits_{\beta=2}^{n-1}\left\langle\nabla_{\by_\beta}\frac{g_{\alpha\omega}}{\chi},
\nabla_{\by_\beta}\chi\right\rangle,
\label{ring2}
\ee
$$
g_{\alpha\omega}\equiv\langle\ba_\alpha,\nabla_{\bp_\omega}\chi\rangle, \ \ \
\ba_\alpha\equiv\hz_1-\varepsilon_{\alpha 1}\hk_\alpha.
$$

To simplify the computations, we will assume that we are beyond the vicinities of zeros
of the function $\chi$. We will only note that the method we use \cite{BL2} allows
to describe the result also when such limitations are missing.

With regard to the expressions
$$
\nabla_{\by_\beta}\frac{g_{\alpha\omega}}{\chi}=
\frac{\nabla_{\by_\beta}g_{\alpha\omega}\chi-g_{\alpha\omega}\nabla_{\by_\beta}\chi}{\chi^2},
$$
$$
\Delta_{\by_\beta}\frac{g_{\alpha\omega}}{\chi}=\frac{1}{\chi}\Delta_{\by_\beta}g_{\alpha\omega}-
2\frac{\langle\nabla_{\by_\beta}\chi,\nabla_{\by_\beta}g_{\alpha\omega}\rangle}{\chi^2}+
2g_{\alpha\omega}\frac{|\nabla_{\by_\beta}\chi|^2}{\chi^3}-
\frac{g_{\alpha\omega}}{\chi^2}\Delta_{\by_\beta}\chi
$$
we will return to the presentation for
$\sigma_\omega^\alpha$  (\ref{ring2})
\be
\sigma_\omega^\alpha=2\langle\bp_\omega,\ba_\alpha\rangle\chi\ +\
\mathop{\sum}\limits_{\beta=2}^{n-1}\left(
\Delta_{\by_\beta}\langle\nabla_{\bp_\omega}\chi,\ba_\alpha\rangle\ -\
\frac{\langle\nabla_{\bp_\omega}\chi,\ba_\alpha\rangle}{\chi}\Delta_{\by_\beta}\chi\right).
\label{eq1-0}
\ee

Remember now that as was noted above in (\ref{chi1}) the function $\chi$ is a "cluster" wave function
 and it depends on a set of variables
$\by_\beta,\ \beta=2,...,n-1$; $\ \bp_\omega,\ \omega=2,...,n-1$.

We will act on the equation (\ref{shr-c}) by the operator $\nabla_{\bp_\omega}$ and multiply the result
by the independent vector $\ba$ in a scalar sense:
\be
-\mathop{\sum}\limits_{\beta=2}^{n-1}\Delta_{\by_\beta}
\langle\nabla_{\bp_\omega}\chi,\ba\rangle\ +\
V_1\langle\nabla_{\bp_\omega}\chi,\ba\rangle\ =\
2\langle\bp_\omega,\ba\rangle\chi\ +\ E_1\langle\nabla_{\bp_\omega}\chi,\ba\rangle.
\label{shr-n}
\ee

The expressions for $V_1$ and $E_1$ were described above in the equation (\ref{claster}).
Substituting the equations (\ref{shr-c}) and (\ref{shr-n}) into the equation (\ref{eq1-0}), we conclude that the following equation is valid
\be
\sigma_\omega^\alpha=0,\ \ \alpha\in\{M_0\},\ \omega=2,3,...,n-1.
\label{result}
\ee
From the equation (\ref{ring1}), thus it follows that
\be
S_\alpha=0,\ \ \alpha\in\{M_0\}.
\label{result1}
\ee

In accordance with the equation (\ref{disca}) we have shown that the full  discrepancy $S$ of the
Schredinger equation decreases at infinity in configuration space as
$O\left(\frac{1}{|\bz_1|^2}\right)$.

\subsection{Extension of the statement about discrepancy decrease velocity to an arbitrary
asymptotic configuration in the  $n$ particles system.}

Now we will return to the general expression for the ansatz (\ref{result-0}). It was noted above that in a limit case when all relative pair Jacobi coordinates go to infinity, the expression (\ref{result-0}) goes into the expression (\ref{res-00}). This case was studied earlier in the work \cite{KL}, where it was shown that the Schredinger equation discrepancy for such an ansatz decreases fast. Note that the "cluster" functions  $\chi_j$ do not contain the big variables $\bz$. Thus the discrepancy computation connected only to these variables, repeats the proof suggested in \cite{KL}.
Note now that according to the constructions offered above, each finite variable $\by$ is an argument of just one "cluster" function  $\chi_j$.
In the sense the computation of the discrepancy connected with finite variables repeats the proof offered above for the case of an arbitrary "cluster" and one well separated particle.

The suggested arguments demonstrate that the Schredinger equation discrepancy for the ansatz (\ref{result-0})describing a leading term of the asymptotics of the $n$ three-dimensional like-charged quantum particles scattering problem solution, decreases faster than the potential at infinity in configuration space uniformly in all angle variables.

\section{Conclusion}

In this work on the basis of the results in the works \cite{BL2},\cite{BL3}, we have offered an expression for an ansatz describing in the terms of formal asymptotic decompositions a leading term of the asymptotics of the $n$ three-dimensional like-charged quantum particles scattering problem solution. We have demonstrated that the Schredinger equation discrepancy for this ansatz decreases faster than the potential at infinity in configuration space uniformly in all angle variables with an exception of, maybe, narrow vicinities  of the forward scattering directions.
In these vicinities the discrepancy decreases faster than the potential too, however, we have not succeeded to obtain exact estimations limited from below.

We will formulate as well {\bf An assumption}.

{\bf An assumption:} {\it The Ansatz of the kind (\ref{result-0}) on the level of formal asymptotic decompositions describes a leading term of the asymptotics of the $n$ three-dimensional quantum particles scattering problem solution for a {\bf broad class} of slowly decreasing pair potentials.}

The definition of this class would be the subject of a separate work.

\end{document}